\documentclass[aip,reprint]{revtex4-1}

\draft % marks overfull lines with a black rule on the right

\usepackage{graphicx}% Include figure files
\usepackage{dcolumn}% Align table columns on decimal point
\usepackage{bm}% bold math
\usepackage{textcomp}
\begin{document}

%\preprint{}

\title{Tunable single-mode laser on thin film lithium niobate} %Title of paper


\author{Xiangmin Liu$ ^{\ast} $}
\affiliation{State Key Laboratory of Advanced Optical Communication Systems and Networks, School of Physics and Astronomy, Shanghai Jiao Tong University, Shanghai 200240, China}

\author{Xiongshuo Yan$ ^{\ast} $}
\affiliation{State Key Laboratory of Advanced Optical Communication Systems and Networks, School of Physics and Astronomy, Shanghai Jiao Tong University, Shanghai 200240, China}

\author{Yi'an Liu}
\affiliation{State Key Laboratory of Advanced Optical Communication Systems and Networks, School of Physics and Astronomy, Shanghai Jiao Tong University, Shanghai 200240, China}

\author{Hao Li}
\affiliation{State Key Laboratory of Advanced Optical Communication Systems and Networks, School of Physics and Astronomy, Shanghai Jiao Tong University, Shanghai 200240, China}

\author{Yuping Chen}
\email{ypchen@sjtu.edu.cn}
\affiliation{State Key Laboratory of Advanced Optical Communication Systems and Networks, School of Physics and Astronomy, Shanghai Jiao Tong University, Shanghai 200240, China}

\author{Xianfeng Chen}
\email{xfchen@sjtu.edu.cn}
\affiliation{State Key Laboratory of Advanced Optical Communication Systems and Networks, School of Physics and Astronomy, Shanghai Jiao Tong University, Shanghai 200240, China}
\affiliation{Shanghai Research Center for Quantum Sciences Shanghai 201315, China}
\affiliation{Jinan Institute of Quantum Technology, Jinan 250101, China}
\affiliation{Collaborative Innovation Center of Light Manipulations and Applications Shandong Normal University, Jinan 250358, China}

%\author{}
%\thanks{}
%\affiliation{}


\date{\today}

\begin{abstract}
Erbium-doped lithium niobate on insulator (LNOI) laser plays an important role in the complete photonic integrated circuits (PICs). Here, we demonstrate an integrated tunable whisper galley single mode laser (WGSML) by making use of a pair of coupled microdisk and microring on LNOI. A 974 nm single-mode pump light can have an excellent resonance in the designed microdisk, which is beneficial to the whisper gallery mode (WGM) laser generation. The WGSML at 1560.40 nm with a maximum 31.4 dB side mode suppression ratio (SMSR) has been achieved. By regulating the temperature, WGSMLs output power increased and the central wavelength can be changed from 1560.30 nm to 1560.40 nm. What's more, 1560.60 nm and 1565.00 nm WGSMLs have been achieved by changing the coupling gap width between microdisk and microring. We can also use the electro-optic effect of LNOI to obtain more accurate adjustable WGSMLs in further research.
\end{abstract}

\pacs{}% insert suggested PACS numbers in braces on next line

\maketitle %\maketitle must follow title, authors, abstract and \pacs

%\section*{Introduction}
	
In recent years, lithium niobate on insulator (LNOI) has become a research focus of photonic integrated circuits (PICs). Due to its excellent material features, such as high electro-optical ($r_{33}=30.8\times 10^{-12}$ m/V) and second-order nonlinear ($d_{33}=-33$ pm/V) coefficients, extraordinary acousto-optic effects, piezoelectric effects, photoelastic effect, and wide transparency window from visible to mid-infrared, LNOI is a promising platform for compact photonic circuits \cite{ref1,ref2}. Plenty of on-chip optical devices have been achieved successfully on LNOI, for instance, electro- and acousto-optical modulators \cite{he2019high,li2020lithium,wang2018integrated}, frequency combs \cite{wang2019monolithic,zhang2019broadband}, nonlinear frequency converter \cite{ye2020sum,liu2019effective,ge2018broadband,jiang2018nonlinear,lin2019broadband,hao2017sum,ref6,ref9} and optomechanical applications  \cite{jiang2020high}, etc. 
However, to achieve complete photonic integrated circuits on LNOI, integrated C-band light sources are essential and the pure lithium niobate (LN) does not have gain characteristics. Inspired by erbium-doped fiber amplifiers, doping rare-earth ions into the LNOI is easily considered a way to overcome this shortcoming. 

Recently, erbium-doped lithium niobate has been developed and a series of researches have been reported including waveguide amplifiers \cite{zhou2021chip,ref13} and microcavity lasers \cite{ref10,ref11,ref12} on $\rm Er^{3+}$-doped LNOI, which exhibit great potential applications in lithium niobate (LN) PICs. Nevertheless, most of the reported lasers based on $\rm Er^{3+}$-doped microdisk are multimodes. The on-chip integrated C-band single-mode laser is critical and need to be developed to improve monochromaticity, stability, and beam quality \cite{chen2021non}. To achieve the on-chip single-mode laser, one can use a microring resonator with a properly designed width to filter out high-order transverse modes. Simultaneously, a single-longitudinal mode laser can be achieved by decreasing the radius of the microring to increase the free spectrum range (FSR) \cite{ref15}. However, the quality (Q) factor of microring reduces with the decrease of the size, which leads to higher loss and lower laser output power. 
There are other approaches to realized single-longitudinal mode output, for example, using gratings to select a particular mode or producing a single-mode laser by breaking PT-symmetry \cite{ref16,ref17,ref18,ref19}. Another more feasible and suitable method for on-chip integration is to fabricate a photonic molecule consisting of two coupled microresonators with different FSRs to achieve mode selection \cite{ref20,ref21}. It is worth noting that research works \cite{gao2021chip,zhang2021integrated,xiao2021single} of LNOI single-mode laser by using two coupled microdisks or two coupled microrings have been published very recently. But two coupled microrings need to use a tunable pump light to achieve an effect laser emission which limits the practical application. What's more, suspended microdisks and coupling light with a tapered fiber are not conducive to integration, which limits its practical application in PICs on LNOI.

Here, we design and fabricate a WGSML on Erbium-doped LNOI based on a photonic molecule which consists of a designed microdisk coupled with a microring resonator. The WGSML with a wavelength of 1560.40 nm is achieved with the 974 nm single-mode pump. The WGSML threshold pump power is about 1.31 mW and the corresponding slope efficiency is $4.41\times 10^{-5}$. By regulating the temperature, the WGSML output power increased and the central wavelength can be changed from 1560.30 nm to 1560.40 nm. A set coupled microdisk and microring with different gap widths are fabricated and 1560.60 nm and 1565.00 nm WGSMLs have been achieved.

Fig. 1(a) shows the structure of the WGSML with a designed microdisk and a microring on a 1$\%$ mol Z-cut $\rm Er^{3+}$-doped LNOI. The LNOI consists of a 600 nm $\rm Er^{3+}$-doped thin film LN, a 2 $\mu$m-silica buffer layer, and a 400 $\mu$m silicon substrate. The Er:LNOI is developed as same as Ref[19]. Fig. 1(b) is the SEM image of the coupled waveguide, the microdisk, and microring with the radii $R_1=150$ $\mu$m and $R_2=165$ $\mu$m, respectively. The microring has a narrow width of 1.2 $\mu$m. The straight bus waveguide top width is 0.6 $\mu$m and the pulley-coupling waveguide top width is 0.95 $\mu$m in order to satisfy the coupling phase-matching condition. The insets show the gap width of the couple regions as that the gap width between the straight coupled waveguide and the microdisk is $g_{1}=500 $ nm,  $g_{20}=360 $ nm between the microdisk and microring, and $g_{3}=770 $ nm in the pulley-coupling region. The main fabrication processes of our designed WGSMLs includes: first we deposit a 600-nm-thick amorphous silicon on the $\rm Er^{3+}$-doped LNOI as a hard etching mask and spin-coat a layer of resist; then the configuration of the coupled microcavities is patterned by electron-beam lithography (EBL) and the mask layer patterns are transferred on the $\rm Er^{3+}$-doped LNOI layer surface via inductively coupled plasma-reactive ion etching (ICP-RIE); finally the remaining mask is removed by wet etching.    

\begin{figure}[htbp]
	\centering
	\includegraphics[width=\linewidth]{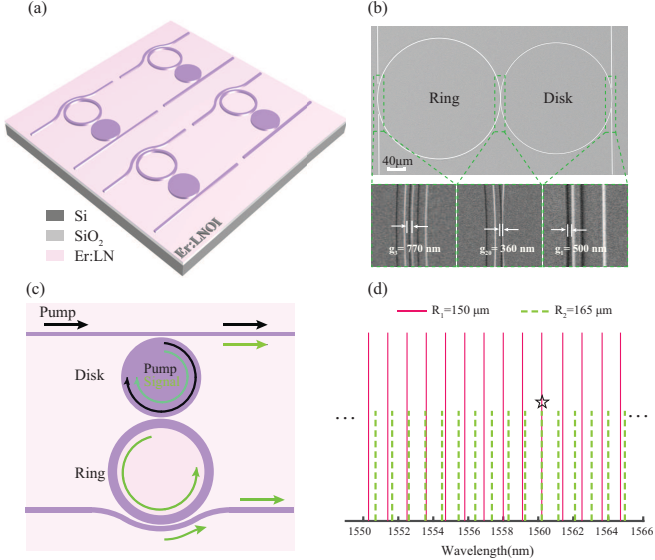}
	\caption{
		(a) Schematic diagram of the photonic molecule on $\rm Er^{3+}$-doped LNOI. (b) The scanning electron microscope (SEM) image of the fabricated sample (top) and the coupling region among straight waveguide, microdisk, microring, and pulley-coupling waveguide (bottom). (c) Principle of the WGSML emission in the microring and microdisk. (d) Schematic diagram of the resonant wavelength of two coupled micro-resonators with different radii. Red solid lines and green dashed lines are numerically calculated resonance wavelengths of the microdisk with a radius of $R_1=150$ $\mu$m and the microring with a radius of $R_2=165$ $\mu$m. The pentagram points to the same wavelength resonants at 1560.20 nm.}
	\label{fig:false-color}
\end{figure}

Fig. 1(c) shows the brief principle of the WGSML emission in the device. we choose the microdisk to generate laser signals. By designing a relatively large radius ($R_1=150$ $\mu$m), we ensure that the pump light can resonate easily in the microdisk without extra adjustment and more pump power can be stored in the
microdisk. As shown in Fig. 1(c), the pump light is coupled into the microdisk through the straight bus waveguide and resonates to produce laser signals. Then the signal light generated by the microdisk is coupled into the microring through the evanescent wave with some certain WGM, which finally couple out of the microring through the pulley-coupling waveguide. 

Fig. 1(d) is the schematic diagram of the resonant wavelength of a coupled microdisk and microring  with different radii. The red solid line and green dashed line are numerically calculated resonance wavelengths of the microdisk and microring. As we know, the mode supported by resonators should satisfy the resonant conditions: $2\pi Rn_{eff}=m\lambda$, where R is the radius of resonators, $n_{eff}$ is the effective refractive index, m is the model index and $\lambda_m$ is the resonant wavelength. The FSR of $R_1$ and  $R_2$ are 1.1 nm and 0.95 nm, respectively. The resonance super-mode of the coupled system exists when the resonances of the microdisk and microring coincide \cite{ref22}, as Fig. 1(d) displayed that only the super-mode with $\lambda=1560.20$ nm is supported by the photonic molecule in the range of 1550 nm to 1565 nm. Therefore, the FSR of the photonic molecule around 1560.20 nm is $FSR\approx \lambda_0^2/{2\pi n_{neff}(R_2-R_1)}=13.7$ nm, which is much larger than microdisk and ring, thus leading to a reduction in the number of longitudinal modes. Facilitated by gain competition and the resonant condition, we can achieve single-mode laser emission by this device.

\begin{figure}[htbp]
	\centering
	\includegraphics[width=\linewidth]{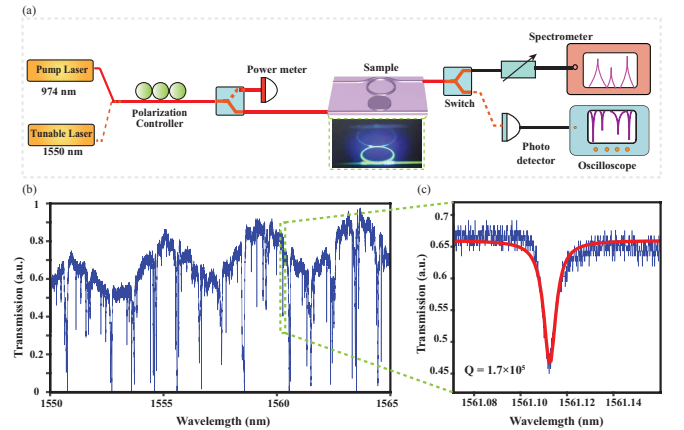}
	\caption{(a) Schematic diagram of the experimental setup. Inset: the optical microscope
		images. (b) Transmission spectrum of the photonic molecule from 1550 nm to 1560 nm with  $ g_{20}=360  $ nm, $ g_{1}=500$ nm and $ g_{3}=770 $ nm. (c) Lorentzian fitting of a measured mode at 1561.11 nm wavelength in the green dashed frame in (b), exhibiting a Q factor of $1.7 \times 10^5$.}
	\label{fig:false-color}
	
\end{figure}
The experimental setup is shown in Fig. 2(a). A 974 nm LD light source (Golight Co., Ltd) is used as the pump source to produce laser emission in the microdisk on the $\rm Er^{3+}$-doped LNOI. The pump light propagates through the polarization controller (PC) and couples into the straight waveguide. The device was placed on a thermalelectric cooler (TEC) to adjust the temperature. Then laser signals are collected and analyzed by the spectrometer to obtain emission spectra. A power meter is connected to the PC to measure the pump power and a tunable continuous-wave laser in C-band (1520-1570 nm) and oscilloscope are used to analyze the transmission spectrum of the device. During the experiment, green up-conversion fluorescence is observed in the device from the photograph of the 974-nm-pumped microdisk, shown in the inset (green dashed frame) of Fig. 2(a). On the one hand, this phenomenon illustrates that the pump light is coupled and localized well in the device. On the other hand, 974-nm-pump light mainly resonants in the microdisk, which avoids the pump light resonating in the microring to produce laser signals of other longitudinal modes. Fig. 2(b) shows the transmission spectrum of the coupled microring and microdisk from 1550 nm to 1565 nm. A Q factor of $1.7\times 10^5$ is evaluated from the Lorentz fitting in Fig. 2(c) for the mode around 1561.11 nm in the green dashed box in Fig. 2(b), which shows the device has a low loss of laser signal and thus a relatively low single-mode threshold pump power.

\begin{figure}[htbp]
	\centering
	\includegraphics[width=\linewidth]{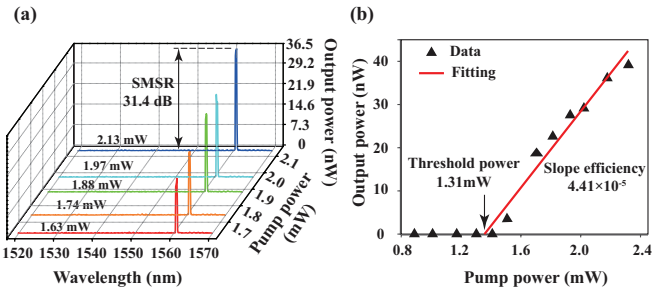}
	\caption{(a) The spectra of the WGSML under different wavelength and pump powers, with $g_{20}=360$ nm and T=25 \textcelsius. The side mode suppression ratio (SMSR) is 31.4 dB. (b) The relationship between the WGSML 1560.40 nm power and the 974 nm pump power.}
	\label{fig:false-color}
\end{figure}

Fig. 3(a) illustrates WGSML emission spectra in the range of 1520 nm to 1570 nm under different pump power of 1.63 mW, 1.74 mW, 1.88 mW, 1.97 mW and 2.13 mW. A WGSML with a wavelength of 1560.40 nm can be obtained stably from the photonic molecules. With the increase of the pump power, the peak power of the laser signal increases and the center wavelength keeps stable. The peak laser power with 2.13 mW pump light is -44.02 dBm and the relevant side mode suppression ratio (SMSR) is 31.40 dB, which demonstrates prominent single-mode characteristics. Fig. 3(b) demonstrates the output laser power as a function of pump power, which indicates that the threshold pump power is 1.31 mW and the slope efficiency is $4.41\times 10^{-5}$. Although the FSR (13.7 nm) of the photonic molecule is smaller than the gain spectrum range of $\rm Er^{3+}$, the coupled system can still achieve a single-mode laser output due to gain competition.

\begin{figure}[htbp]
	\centering
	\includegraphics[width=\linewidth]{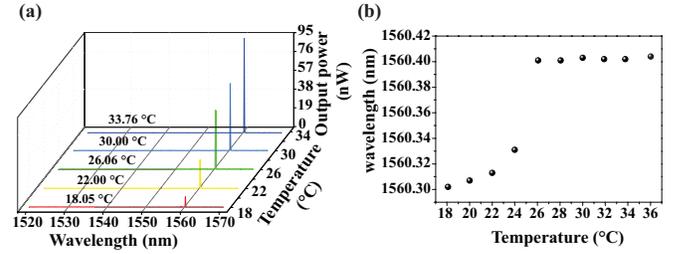}
	\caption{(a) The spectra of the WGSML under different wavelength and temperature, with $g_{20}=360$ nm and a fixed pump power of 1.88 mW.  (b) The relationship between the WGSML central wavelength and  temperature.}
	\label{fig:false-color}
\end{figure}
%\par
Fig. 4(a) shows WGSMLs emission spectra with a fixed pump power of 1.88 mW under different temperatures of 18.05 \textcelsius, 22.00 \textcelsius , 26.06 \textcelsius, 30.00\textcelsius, 33.76\textcelsius, respectively. We found that the WGSML output power increases with  the temperature, which may be caused by more pump light resonating to produce signal light under the refractive index changing with the temperature. What's more, we also found that the WGSML central wavelength changing with the temperature, shown in Fig. 4(b), which should be caused by the resonance condition changing due to the thermal effect of the LNOI.
The WGSML central wavelength shows the increasing tendency with the temperature from 1560.30 nm to 1560.40 nm, then maintains certain stability in the range from 26 \textcelsius  to 34 \textcelsius. It is worth stating that more accurate laser wavelength regulation can be achieved by using the electro–optic effect of LNOI \cite{guarino2007electro}.

\begin{figure}[htbp]
	\centering
	\includegraphics[width=\linewidth]{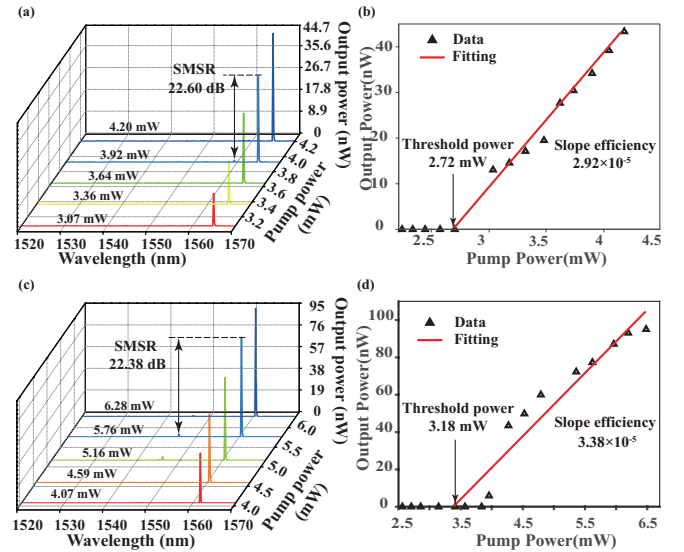}
	\caption{(a), (c) are the spectra of the WGSML under different wavelength and pump powers, with T=25\textcelsius, $ g_{21}=410$ nm and $g_{22}=460 $ nm, respectively. (b), (d) are the relationship between the 1565.00 nm and 1560.60 nm WGSMLs power and the 974-nm pump power, respectively. $ g_{1} $ and $ g_{3} $ are the same as before. }	
\end{figure}

The effect of gap width between microdisk and microring on WGSMLs is also analyzed in our experiment. A set of coupled microring and microdisk with a different gap width of $ g_{21}=410$ nm and $g_{22}=460 $ nm are fabricated and measured. Fig. 5(a) and (b) show the WGSML emission spectra and output WGSMLs power versus different pump power with $ g_{21}=410$ nm . The emitted WGSML is at 1565.00 nm and has an SMSR of 22.60 dB, shown in Fig. 5(a). The corresponding threshold pump power is 2.72 mW and the slope efficiency is $2.92 \times 10^{-5}$. Fig. 5(c) and (d) illustrate the laser spectra and the relation between the pump power and the output power with $g_{22}=460 $ nm. Its threshold pump power and slope efficiency are 3.18 mW and $3.38\times 10^{-5}$. Comparing with the laser at 1560.40 nm of $g_{20}=360 $ nm, the laser wavelength of $g_{22}=460 $ nm changes larger to 1560.60 nm. 
What's more, the SMSR of the single-mode laser with a 460 nm gap is 22.38 dB, which still maintains a relatively satisfactory single mode property.

In addition, comparing the WGSML threshold pump power with $g_{20}=360 $ nm,  $g_{21}=410 $ nm, and $g_{22}=460 $ nm, the increased gap width leads to a decrease in the coupling efficiency of the microdisk and microring. This behavior causes more energy to lay in those WGMs which merely exist in microdisk and do not resonate in microring. Thus WGSML threshold power increases with the gap width. Similarly, the WGSML slope efficiency with $ g_{22}=460 $ nm  ($3.38 \times 10^{-5}$) is reduced compared to laser with  $ g_{20}=360 $ nm ($4.41 \times 10^{-5}$), which should be caused by the decrease of the coupling coefficient. However, we find an abnormal mode change of the WGSML with $g_{21}=410 $ nm, which leads to the slope efficiency ($2.92\times 10^{-5}$) less than the other two single mode lasers slope efficiency with  $ g_{20}=360 $ nm and  $ g_{22}=460 $ nm. The mode change may be caused by the gain competition and the FSR of the microdisk and microring is fairly small due to their large radii and dense modes distribution results that the modes near the resonant super-mode can still exist in the device \cite{ref23}. Our numerical simulation results also explain the mode change from the same perspective. Calculated resonant wavelengths in Fig. 1(d) display that the microdisk support mode is 1564.80 nm while the microring supports 1565.00 nm. This tiny deviation may enable signal light at 1565.00 nm to resonate in the coupling system. What's more, with the increase of pump power, more pump light may also couple into the microring. Thus the WGSML at 1565.00 nm in the microring is emitted after the gain competition, which is consistent with our calculation that the mode at 1565.00 nm is supported by microring. More details still need to be further studied.

In conclusion, an integrated WGSML has been fabricated on $\rm Er^{3+}$-doped LNOI. Facilitated by a coupled microdisk and microring with different radii, WGSML at the wavelength of 1560.40 nm can be stably emitted with a threshold pump power of 1.31 mW and slope efficiency of $4.41\times 10^{-5}$. A maximum 31.4 dB SMSR is achieved. The WGSML wavelength and output power changed by adjusting the temperature are observed. The effect of gap width between microdisk and microring on WGSML is also analyzed. With the integrated C-band single-mode laser on Erbium-doped LNOI, a series of on-chip optical devices and applications can be developed and show great potential in LNOI PICs.

This work was supported by the National Key R $\&$ D Program of China (Grant Nos. 2019YFB2203501, 2017YFA0303701), the National Natural Science Foundation of China (Grant Nos. 12134009, 91950107, 11734011), Shanghai Municipal Science and Technology Major Project
(2019SHZDZX01-ZX06), and SJTU No. 21X010200828.
 
[$ ^{\ast} $] These authors contributed equally to this Letter.		

\bibliography{References}

\begin{thebibliography}{35}%
\makeatletter
\providecommand \@ifxundefined [1]{%
 \@ifx{#1\undefined}
}%
\providecommand \@ifnum [1]{%
 \ifnum #1\expandafter \@firstoftwo
 \else \expandafter \@secondoftwo
 \fi
}%
\providecommand \@ifx [1]{%
 \ifx #1\expandafter \@firstoftwo
 \else \expandafter \@secondoftwo
 \fi
}%
\providecommand \natexlab [1]{#1}%
\providecommand \enquote  [1]{``#1''}%
\providecommand \bibnamefont  [1]{#1}%
\providecommand \bibfnamefont [1]{#1}%
\providecommand \citenamefont [1]{#1}%
\providecommand \href@noop [0]{\@secondoftwo}%
\providecommand \href [0]{\begingroup \@sanitize@url \@href}%
\providecommand \@href[1]{\@@startlink{#1}\@@href}%
\providecommand \@@href[1]{\endgroup#1\@@endlink}%
\providecommand \@sanitize@url [0]{\catcode `\\12\catcode `\$12\catcode
  `\&12\catcode `\#12\catcode `\^12\catcode `\_12\catcode `\%12\relax}%
\providecommand \@@startlink[1]{}%
\providecommand \@@endlink[0]{}%
\providecommand \url  [0]{\begingroup\@sanitize@url \@url }%
\providecommand \@url [1]{\endgroup\@href {#1}{\urlprefix }}%
\providecommand \urlprefix  [0]{URL }%
\providecommand \Eprint [0]{\href }%
\providecommand \doibase [0]{http://dx.doi.org/}%
\providecommand \selectlanguage [0]{\@gobble}%
\providecommand \bibinfo  [0]{\@secondoftwo}%
\providecommand \bibfield  [0]{\@secondoftwo}%
\providecommand \translation [1]{[#1]}%
\providecommand \BibitemOpen [0]{}%
\providecommand \bibitemStop [0]{}%
\providecommand \bibitemNoStop [0]{.\EOS\space}%
\providecommand \EOS [0]{\spacefactor3000\relax}%
\providecommand \BibitemShut  [1]{\csname bibitem#1\endcsname}%
\let\auto@bib@innerbib\@empty
%</preamble>
\bibitem [{\citenamefont {{Turner}}(1966)}]{ref1}%
  \BibitemOpen
  \bibfield  {author} {\bibinfo {author} {\bibfnamefont {E.~H.}\ \bibnamefont
  {{Turner}}},\ }\bibfield  {title} {\enquote {\bibinfo {title}
  {{High-Frequency Electro-Optic Coefficients of Lithium Niobate}},}\ }\href
  {\doibase 10.1063/1.1754449} {\bibfield  {journal} {\bibinfo  {journal}
  {Applied Physics Letters}\ }\textbf {\bibinfo {volume} {8}},\ \bibinfo
  {pages} {303--304} (\bibinfo {year} {1966})}\BibitemShut {NoStop}%
\bibitem [{\citenamefont {Boes}\ \emph {et~al.}(2018)\citenamefont {Boes},
  \citenamefont {Corcoran}, \citenamefont {Chang}, \citenamefont {Bowers},\
  and\ \citenamefont {Mitchell}}]{ref2}%
  \BibitemOpen
  \bibfield  {author} {\bibinfo {author} {\bibfnamefont {A.}~\bibnamefont
  {Boes}}, \bibinfo {author} {\bibfnamefont {B.}~\bibnamefont {Corcoran}},
  \bibinfo {author} {\bibfnamefont {L.}~\bibnamefont {Chang}}, \bibinfo
  {author} {\bibfnamefont {J.}~\bibnamefont {Bowers}}, \ and\ \bibinfo {author}
  {\bibfnamefont {A.}~\bibnamefont {Mitchell}},\ }\bibfield  {title} {\enquote
  {\bibinfo {title} {Status and potential of lithium niobate on insulator
  (lnoi) for photonic integrated circuits},}\ }\href {\doibase
  https://doi.org/10.1002/lpor.201700256} {\bibfield  {journal} {\bibinfo
  {journal} {Laser \& Photonics Reviews}\ }\textbf {\bibinfo {volume} {12}},\
  \bibinfo {pages} {1700256} (\bibinfo {year} {2018})},\ \Eprint
  {http://arxiv.org/abs/https://onlinelibrary.wiley.com/doi/pdf/10.1002/lpor.201700256}
  {https://onlinelibrary.wiley.com/doi/pdf/10.1002/lpor.201700256} \BibitemShut
  {NoStop}%
\bibitem [{\citenamefont {He}\ \emph {et~al.}(2019)\citenamefont {He},
  \citenamefont {Xu}, \citenamefont {Ren}, \citenamefont {Jian}, \citenamefont
  {Ruan}, \citenamefont {Xu}, \citenamefont {Gao}, \citenamefont {Sun},
  \citenamefont {Wen}, \citenamefont {Zhou} \emph {et~al.}}]{he2019high}%
  \BibitemOpen
  \bibfield  {author} {\bibinfo {author} {\bibfnamefont {M.}~\bibnamefont
  {He}}, \bibinfo {author} {\bibfnamefont {M.}~\bibnamefont {Xu}}, \bibinfo
  {author} {\bibfnamefont {Y.}~\bibnamefont {Ren}}, \bibinfo {author}
  {\bibfnamefont {J.}~\bibnamefont {Jian}}, \bibinfo {author} {\bibfnamefont
  {Z.}~\bibnamefont {Ruan}}, \bibinfo {author} {\bibfnamefont {Y.}~\bibnamefont
  {Xu}}, \bibinfo {author} {\bibfnamefont {S.}~\bibnamefont {Gao}}, \bibinfo
  {author} {\bibfnamefont {S.}~\bibnamefont {Sun}}, \bibinfo {author}
  {\bibfnamefont {X.}~\bibnamefont {Wen}}, \bibinfo {author} {\bibfnamefont
  {L.}~\bibnamefont {Zhou}},  \emph {et~al.},\ }\bibfield  {title} {\enquote
  {\bibinfo {title} {High-performance hybrid silicon and lithium niobate
  mach--zehnder modulators for 100 gbit s- 1 and beyond},}\ }\href@noop {}
  {\bibfield  {journal} {\bibinfo  {journal} {Nature Photonics}\ }\textbf
  {\bibinfo {volume} {13}},\ \bibinfo {pages} {359--364} (\bibinfo {year}
  {2019})}\BibitemShut {NoStop}%
\bibitem [{\citenamefont {Li}\ \emph {et~al.}(2020)\citenamefont {Li},
  \citenamefont {Ling}, \citenamefont {He}, \citenamefont {Javid},
  \citenamefont {Xue},\ and\ \citenamefont {Lin}}]{li2020lithium}%
  \BibitemOpen
  \bibfield  {author} {\bibinfo {author} {\bibfnamefont {M.}~\bibnamefont
  {Li}}, \bibinfo {author} {\bibfnamefont {J.}~\bibnamefont {Ling}}, \bibinfo
  {author} {\bibfnamefont {Y.}~\bibnamefont {He}}, \bibinfo {author}
  {\bibfnamefont {U.~A.}\ \bibnamefont {Javid}}, \bibinfo {author}
  {\bibfnamefont {S.}~\bibnamefont {Xue}}, \ and\ \bibinfo {author}
  {\bibfnamefont {Q.}~\bibnamefont {Lin}},\ }\bibfield  {title} {\enquote
  {\bibinfo {title} {Lithium niobate photonic-crystal electro-optic
  modulator},}\ }\href@noop {} {\bibfield  {journal} {\bibinfo  {journal}
  {Nature Communications}\ }\textbf {\bibinfo {volume} {11}},\ \bibinfo {pages}
  {1--8} (\bibinfo {year} {2020})}\BibitemShut {NoStop}%
\bibitem [{\citenamefont {Wang}\ \emph {et~al.}(2018)\citenamefont {Wang},
  \citenamefont {Zhang}, \citenamefont {Chen}, \citenamefont {Bertrand},
  \citenamefont {Shams-Ansari}, \citenamefont {Chandrasekhar}, \citenamefont
  {Winzer},\ and\ \citenamefont {Lon{\v{c}}ar}}]{wang2018integrated}%
  \BibitemOpen
  \bibfield  {author} {\bibinfo {author} {\bibfnamefont {C.}~\bibnamefont
  {Wang}}, \bibinfo {author} {\bibfnamefont {M.}~\bibnamefont {Zhang}},
  \bibinfo {author} {\bibfnamefont {X.}~\bibnamefont {Chen}}, \bibinfo {author}
  {\bibfnamefont {M.}~\bibnamefont {Bertrand}}, \bibinfo {author}
  {\bibfnamefont {A.}~\bibnamefont {Shams-Ansari}}, \bibinfo {author}
  {\bibfnamefont {S.}~\bibnamefont {Chandrasekhar}}, \bibinfo {author}
  {\bibfnamefont {P.}~\bibnamefont {Winzer}}, \ and\ \bibinfo {author}
  {\bibfnamefont {M.}~\bibnamefont {Lon{\v{c}}ar}},\ }\bibfield  {title}
  {\enquote {\bibinfo {title} {Integrated lithium niobate electro-optic
  modulators operating at cmos-compatible voltages},}\ }\href@noop {}
  {\bibfield  {journal} {\bibinfo  {journal} {Nature}\ }\textbf {\bibinfo
  {volume} {562}},\ \bibinfo {pages} {101--104} (\bibinfo {year}
  {2018})}\BibitemShut {NoStop}%
\bibitem [{\citenamefont {Wang}\ \emph {et~al.}(2019)\citenamefont {Wang},
  \citenamefont {Zhang}, \citenamefont {Yu}, \citenamefont {Zhu}, \citenamefont
  {Hu},\ and\ \citenamefont {Loncar}}]{wang2019monolithic}%
  \BibitemOpen
  \bibfield  {author} {\bibinfo {author} {\bibfnamefont {C.}~\bibnamefont
  {Wang}}, \bibinfo {author} {\bibfnamefont {M.}~\bibnamefont {Zhang}},
  \bibinfo {author} {\bibfnamefont {M.}~\bibnamefont {Yu}}, \bibinfo {author}
  {\bibfnamefont {R.}~\bibnamefont {Zhu}}, \bibinfo {author} {\bibfnamefont
  {H.}~\bibnamefont {Hu}}, \ and\ \bibinfo {author} {\bibfnamefont
  {M.}~\bibnamefont {Loncar}},\ }\bibfield  {title} {\enquote {\bibinfo {title}
  {Monolithic lithium niobate photonic circuits for kerr frequency comb
  generation and modulation},}\ }\href@noop {} {\bibfield  {journal} {\bibinfo
  {journal} {Nature Communications}\ }\textbf {\bibinfo {volume} {10}},\
  \bibinfo {pages} {1--6} (\bibinfo {year} {2019})}\BibitemShut {NoStop}%
\bibitem [{\citenamefont {Zhang}\ \emph {et~al.}(2019)\citenamefont {Zhang},
  \citenamefont {Buscaino}, \citenamefont {Wang}, \citenamefont {Shams-Ansari},
  \citenamefont {Reimer}, \citenamefont {Zhu}, \citenamefont {Kahn},\ and\
  \citenamefont {Lon{\v{c}}ar}}]{zhang2019broadband}%
  \BibitemOpen
  \bibfield  {author} {\bibinfo {author} {\bibfnamefont {M.}~\bibnamefont
  {Zhang}}, \bibinfo {author} {\bibfnamefont {B.}~\bibnamefont {Buscaino}},
  \bibinfo {author} {\bibfnamefont {C.}~\bibnamefont {Wang}}, \bibinfo {author}
  {\bibfnamefont {A.}~\bibnamefont {Shams-Ansari}}, \bibinfo {author}
  {\bibfnamefont {C.}~\bibnamefont {Reimer}}, \bibinfo {author} {\bibfnamefont
  {R.}~\bibnamefont {Zhu}}, \bibinfo {author} {\bibfnamefont {J.~M.}\
  \bibnamefont {Kahn}}, \ and\ \bibinfo {author} {\bibfnamefont
  {M.}~\bibnamefont {Lon{\v{c}}ar}},\ }\bibfield  {title} {\enquote {\bibinfo
  {title} {Broadband electro-optic frequency comb generation in a lithium
  niobate microring resonator},}\ }\href@noop {} {\bibfield  {journal}
  {\bibinfo  {journal} {Nature}\ }\textbf {\bibinfo {volume} {568}},\ \bibinfo
  {pages} {373--377} (\bibinfo {year} {2019})}\BibitemShut {NoStop}%
\bibitem [{\citenamefont {Ye}\ \emph {et~al.}(2020)\citenamefont {Ye},
  \citenamefont {Liu}, \citenamefont {Chen}, \citenamefont {Zheng},\ and\
  \citenamefont {Chen}}]{ye2020sum}%
  \BibitemOpen
  \bibfield  {author} {\bibinfo {author} {\bibfnamefont {X.}~\bibnamefont
  {Ye}}, \bibinfo {author} {\bibfnamefont {S.}~\bibnamefont {Liu}}, \bibinfo
  {author} {\bibfnamefont {Y.}~\bibnamefont {Chen}}, \bibinfo {author}
  {\bibfnamefont {Y.}~\bibnamefont {Zheng}}, \ and\ \bibinfo {author}
  {\bibfnamefont {X.}~\bibnamefont {Chen}},\ }\bibfield  {title} {\enquote
  {\bibinfo {title} {Sum-frequency generation in lithium-niobate-on-insulator
  microdisk via modal phase matching},}\ }\href@noop {} {\bibfield  {journal}
  {\bibinfo  {journal} {Optics Letters}\ }\textbf {\bibinfo {volume} {45}},\
  \bibinfo {pages} {523--526} (\bibinfo {year} {2020})}\BibitemShut {NoStop}%
\bibitem [{\citenamefont {Liu}\ \emph {et~al.}(2019)\citenamefont {Liu},
  \citenamefont {Zheng}, \citenamefont {Fang}, \citenamefont {Ye},
  \citenamefont {Cheng},\ and\ \citenamefont {Chen}}]{liu2019effective}%
  \BibitemOpen
  \bibfield  {author} {\bibinfo {author} {\bibfnamefont {S.}~\bibnamefont
  {Liu}}, \bibinfo {author} {\bibfnamefont {Y.}~\bibnamefont {Zheng}}, \bibinfo
  {author} {\bibfnamefont {Z.}~\bibnamefont {Fang}}, \bibinfo {author}
  {\bibfnamefont {X.}~\bibnamefont {Ye}}, \bibinfo {author} {\bibfnamefont
  {Y.}~\bibnamefont {Cheng}}, \ and\ \bibinfo {author} {\bibfnamefont
  {X.}~\bibnamefont {Chen}},\ }\bibfield  {title} {\enquote {\bibinfo {title}
  {Effective four-wave mixing in the lithium niobate on insulator microdisk by
  cascading quadratic processes},}\ }\href@noop {} {\bibfield  {journal}
  {\bibinfo  {journal} {Optics Letters}\ }\textbf {\bibinfo {volume} {44}},\
  \bibinfo {pages} {1456--1459} (\bibinfo {year} {2019})}\BibitemShut {NoStop}%
\bibitem [{\citenamefont {Ge}\ \emph {et~al.}(2018)\citenamefont {Ge},
  \citenamefont {Chen}, \citenamefont {Jiang}, \citenamefont {Li},
  \citenamefont {Zhu}, \citenamefont {Chen} \emph {et~al.}}]{ge2018broadband}%
  \BibitemOpen
  \bibfield  {author} {\bibinfo {author} {\bibfnamefont {L.}~\bibnamefont
  {Ge}}, \bibinfo {author} {\bibfnamefont {Y.}~\bibnamefont {Chen}}, \bibinfo
  {author} {\bibfnamefont {H.}~\bibnamefont {Jiang}}, \bibinfo {author}
  {\bibfnamefont {G.}~\bibnamefont {Li}}, \bibinfo {author} {\bibfnamefont
  {B.}~\bibnamefont {Zhu}}, \bibinfo {author} {\bibfnamefont {X.}~\bibnamefont
  {Chen}},  \emph {et~al.},\ }\bibfield  {title} {\enquote {\bibinfo {title}
  {Broadband quasi-phase matching in a mgo: Ppln thin film},}\ }\href@noop {}
  {\bibfield  {journal} {\bibinfo  {journal} {Photonics Research}\ }\textbf
  {\bibinfo {volume} {6}},\ \bibinfo {pages} {954--958} (\bibinfo {year}
  {2018})}\BibitemShut {NoStop}%
\bibitem [{\citenamefont {Jiang}\ \emph {et~al.}(2018)\citenamefont {Jiang},
  \citenamefont {Liang}, \citenamefont {Luo}, \citenamefont {Chen},
  \citenamefont {Chen},\ and\ \citenamefont {Lin}}]{jiang2018nonlinear}%
  \BibitemOpen
  \bibfield  {author} {\bibinfo {author} {\bibfnamefont {H.}~\bibnamefont
  {Jiang}}, \bibinfo {author} {\bibfnamefont {H.}~\bibnamefont {Liang}},
  \bibinfo {author} {\bibfnamefont {R.}~\bibnamefont {Luo}}, \bibinfo {author}
  {\bibfnamefont {X.}~\bibnamefont {Chen}}, \bibinfo {author} {\bibfnamefont
  {Y.}~\bibnamefont {Chen}}, \ and\ \bibinfo {author} {\bibfnamefont
  {Q.}~\bibnamefont {Lin}},\ }\bibfield  {title} {\enquote {\bibinfo {title}
  {Nonlinear frequency conversion in one dimensional lithium niobate photonic
  crystal nanocavities},}\ }\href@noop {} {\bibfield  {journal} {\bibinfo
  {journal} {Applied Physics Letters}\ }\textbf {\bibinfo {volume} {113}},\
  \bibinfo {pages} {021104} (\bibinfo {year} {2018})}\BibitemShut {NoStop}%
\bibitem [{\citenamefont {Lin}\ \emph {et~al.}(2019)\citenamefont {Lin},
  \citenamefont {Yao}, \citenamefont {Hao}, \citenamefont {Zhang},
  \citenamefont {Mao}, \citenamefont {Wang}, \citenamefont {Chu}, \citenamefont
  {Wu}, \citenamefont {Fang}, \citenamefont {Qiao} \emph
  {et~al.}}]{lin2019broadband}%
  \BibitemOpen
  \bibfield  {author} {\bibinfo {author} {\bibfnamefont {J.}~\bibnamefont
  {Lin}}, \bibinfo {author} {\bibfnamefont {N.}~\bibnamefont {Yao}}, \bibinfo
  {author} {\bibfnamefont {Z.}~\bibnamefont {Hao}}, \bibinfo {author}
  {\bibfnamefont {J.}~\bibnamefont {Zhang}}, \bibinfo {author} {\bibfnamefont
  {W.}~\bibnamefont {Mao}}, \bibinfo {author} {\bibfnamefont {M.}~\bibnamefont
  {Wang}}, \bibinfo {author} {\bibfnamefont {W.}~\bibnamefont {Chu}}, \bibinfo
  {author} {\bibfnamefont {R.}~\bibnamefont {Wu}}, \bibinfo {author}
  {\bibfnamefont {Z.}~\bibnamefont {Fang}}, \bibinfo {author} {\bibfnamefont
  {L.}~\bibnamefont {Qiao}},  \emph {et~al.},\ }\bibfield  {title} {\enquote
  {\bibinfo {title} {Broadband quasi-phase-matched harmonic generation in an
  on-chip monocrystalline lithium niobate microdisk resonator},}\ }\href@noop
  {} {\bibfield  {journal} {\bibinfo  {journal} {Physical Review Letters}\
  }\textbf {\bibinfo {volume} {122}},\ \bibinfo {pages} {173903} (\bibinfo
  {year} {2019})}\BibitemShut {NoStop}%
\bibitem [{\citenamefont {Hao}\ \emph {et~al.}(2017)\citenamefont {Hao},
  \citenamefont {Wang}, \citenamefont {Ma}, \citenamefont {Mao}, \citenamefont
  {Bo}, \citenamefont {Gao}, \citenamefont {Zhang},\ and\ \citenamefont
  {Xu}}]{hao2017sum}%
  \BibitemOpen
  \bibfield  {author} {\bibinfo {author} {\bibfnamefont {Z.}~\bibnamefont
  {Hao}}, \bibinfo {author} {\bibfnamefont {J.}~\bibnamefont {Wang}}, \bibinfo
  {author} {\bibfnamefont {S.}~\bibnamefont {Ma}}, \bibinfo {author}
  {\bibfnamefont {W.}~\bibnamefont {Mao}}, \bibinfo {author} {\bibfnamefont
  {F.}~\bibnamefont {Bo}}, \bibinfo {author} {\bibfnamefont {F.}~\bibnamefont
  {Gao}}, \bibinfo {author} {\bibfnamefont {G.}~\bibnamefont {Zhang}}, \ and\
  \bibinfo {author} {\bibfnamefont {J.}~\bibnamefont {Xu}},\ }\bibfield
  {title} {\enquote {\bibinfo {title} {Sum-frequency generation in on-chip
  lithium niobate microdisk resonators},}\ }\href@noop {} {\bibfield  {journal}
  {\bibinfo  {journal} {Photonics Research}\ }\textbf {\bibinfo {volume} {5}},\
  \bibinfo {pages} {623--628} (\bibinfo {year} {2017})}\BibitemShut {NoStop}%
\bibitem [{\citenamefont {Shao}\ \emph {et~al.}(2019)\citenamefont {Shao},
  \citenamefont {Yu}, \citenamefont {Maity}, \citenamefont {Sinclair},
  \citenamefont {Zheng}, \citenamefont {Chia}, \citenamefont {Shams-Ansari},
  \citenamefont {Wang}, \citenamefont {Zhang}, \citenamefont {Lai} \emph
  {et~al.}}]{ref6}%
  \BibitemOpen
  \bibfield  {author} {\bibinfo {author} {\bibfnamefont {L.}~\bibnamefont
  {Shao}}, \bibinfo {author} {\bibfnamefont {M.}~\bibnamefont {Yu}}, \bibinfo
  {author} {\bibfnamefont {S.}~\bibnamefont {Maity}}, \bibinfo {author}
  {\bibfnamefont {N.}~\bibnamefont {Sinclair}}, \bibinfo {author}
  {\bibfnamefont {L.}~\bibnamefont {Zheng}}, \bibinfo {author} {\bibfnamefont
  {C.}~\bibnamefont {Chia}}, \bibinfo {author} {\bibfnamefont {A.}~\bibnamefont
  {Shams-Ansari}}, \bibinfo {author} {\bibfnamefont {C.}~\bibnamefont {Wang}},
  \bibinfo {author} {\bibfnamefont {M.}~\bibnamefont {Zhang}}, \bibinfo
  {author} {\bibfnamefont {K.}~\bibnamefont {Lai}},  \emph {et~al.},\
  }\bibfield  {title} {\enquote {\bibinfo {title} {Microwave-to-optical
  conversion using lithium niobate thin-film acoustic resonators},}\
  }\href@noop {} {\bibfield  {journal} {\bibinfo  {journal} {Optica}\ }\textbf
  {\bibinfo {volume} {6}},\ \bibinfo {pages} {1498--1505} (\bibinfo {year}
  {2019})}\BibitemShut {NoStop}%
\bibitem [{\citenamefont {Lin}\ \emph {et~al.}(2020)\citenamefont {Lin},
  \citenamefont {Bo}, \citenamefont {Cheng},\ and\ \citenamefont {Xu}}]{ref9}%
  \BibitemOpen
  \bibfield  {author} {\bibinfo {author} {\bibfnamefont {J.}~\bibnamefont
  {Lin}}, \bibinfo {author} {\bibfnamefont {F.}~\bibnamefont {Bo}}, \bibinfo
  {author} {\bibfnamefont {Y.}~\bibnamefont {Cheng}}, \ and\ \bibinfo {author}
  {\bibfnamefont {J.}~\bibnamefont {Xu}},\ }\bibfield  {title} {\enquote
  {\bibinfo {title} {Advances in on-chip photonic devices based on lithium
  niobate on insulator},}\ }\href@noop {} {\bibfield  {journal} {\bibinfo
  {journal} {Photonics Research}\ }\textbf {\bibinfo {volume} {8}},\ \bibinfo
  {pages} {1910--1936} (\bibinfo {year} {2020})}\BibitemShut {NoStop}%
\bibitem [{\citenamefont {Jiang}\ \emph {et~al.}(2020)\citenamefont {Jiang},
  \citenamefont {Yan}, \citenamefont {Liang}, \citenamefont {Luo},
  \citenamefont {Chen}, \citenamefont {Chen},\ and\ \citenamefont
  {Lin}}]{jiang2020high}%
  \BibitemOpen
  \bibfield  {author} {\bibinfo {author} {\bibfnamefont {H.}~\bibnamefont
  {Jiang}}, \bibinfo {author} {\bibfnamefont {X.}~\bibnamefont {Yan}}, \bibinfo
  {author} {\bibfnamefont {H.}~\bibnamefont {Liang}}, \bibinfo {author}
  {\bibfnamefont {R.}~\bibnamefont {Luo}}, \bibinfo {author} {\bibfnamefont
  {X.}~\bibnamefont {Chen}}, \bibinfo {author} {\bibfnamefont {Y.}~\bibnamefont
  {Chen}}, \ and\ \bibinfo {author} {\bibfnamefont {Q.}~\bibnamefont {Lin}},\
  }\bibfield  {title} {\enquote {\bibinfo {title} {High harmonic optomechanical
  oscillations in the lithium niobate photonic crystal nanocavity},}\
  }\href@noop {} {\bibfield  {journal} {\bibinfo  {journal} {Applied Physics
  Letters}\ }\textbf {\bibinfo {volume} {117}},\ \bibinfo {pages} {081102}
  (\bibinfo {year} {2020})}\BibitemShut {NoStop}%
\bibitem [{\citenamefont {Zhou}\ \emph {et~al.}(2021)\citenamefont {Zhou},
  \citenamefont {Liang}, \citenamefont {Liu}, \citenamefont {Chu},
  \citenamefont {Zhang}, \citenamefont {Yin}, \citenamefont {Fang},
  \citenamefont {Wu}, \citenamefont {Zhang}, \citenamefont {Chen} \emph
  {et~al.}}]{zhou2021chip}%
  \BibitemOpen
  \bibfield  {author} {\bibinfo {author} {\bibfnamefont {J.}~\bibnamefont
  {Zhou}}, \bibinfo {author} {\bibfnamefont {Y.}~\bibnamefont {Liang}},
  \bibinfo {author} {\bibfnamefont {Z.}~\bibnamefont {Liu}}, \bibinfo {author}
  {\bibfnamefont {W.}~\bibnamefont {Chu}}, \bibinfo {author} {\bibfnamefont
  {H.}~\bibnamefont {Zhang}}, \bibinfo {author} {\bibfnamefont
  {D.}~\bibnamefont {Yin}}, \bibinfo {author} {\bibfnamefont {Z.}~\bibnamefont
  {Fang}}, \bibinfo {author} {\bibfnamefont {R.}~\bibnamefont {Wu}}, \bibinfo
  {author} {\bibfnamefont {J.}~\bibnamefont {Zhang}}, \bibinfo {author}
  {\bibfnamefont {W.}~\bibnamefont {Chen}},  \emph {et~al.},\ }\bibfield
  {title} {\enquote {\bibinfo {title} {On-chip integrated waveguide amplifiers
  on erbium-doped thin film lithium niobate on insulator},}\ }\href@noop {}
  {\bibfield  {journal} {\bibinfo  {journal} {Laser \& Photonics Reviews}\ }
  (\bibinfo {year} {2021})}\BibitemShut {NoStop}%
\bibitem [{\citenamefont {Chen}\ \emph
  {et~al.}(2021{\natexlab{a}})\citenamefont {Chen}, \citenamefont {Xu},
  \citenamefont {Zhang}, \citenamefont {Wong}, \citenamefont {Zhang},
  \citenamefont {Pun},\ and\ \citenamefont {Wang}}]{ref13}%
  \BibitemOpen
  \bibfield  {author} {\bibinfo {author} {\bibfnamefont {Z.}~\bibnamefont
  {Chen}}, \bibinfo {author} {\bibfnamefont {Q.}~\bibnamefont {Xu}}, \bibinfo
  {author} {\bibfnamefont {K.}~\bibnamefont {Zhang}}, \bibinfo {author}
  {\bibfnamefont {W.-H.}\ \bibnamefont {Wong}}, \bibinfo {author}
  {\bibfnamefont {D.-L.}\ \bibnamefont {Zhang}}, \bibinfo {author}
  {\bibfnamefont {E.~Y.-B.}\ \bibnamefont {Pun}}, \ and\ \bibinfo {author}
  {\bibfnamefont {C.}~\bibnamefont {Wang}},\ }\bibfield  {title} {\enquote
  {\bibinfo {title} {Efficient erbium-doped thin-film lithium niobate waveguide
  amplifiers},}\ }\href@noop {} {\bibfield  {journal} {\bibinfo  {journal}
  {Optics Letters}\ }\textbf {\bibinfo {volume} {46}},\ \bibinfo {pages}
  {1161--1164} (\bibinfo {year} {2021}{\natexlab{a}})}\BibitemShut {NoStop}%
\bibitem [{\citenamefont {Liu}\ \emph {et~al.}(2021)\citenamefont {Liu},
  \citenamefont {Yan}, \citenamefont {Wu}, \citenamefont {Zhu}, \citenamefont
  {Chen},\ and\ \citenamefont {Chen}}]{ref10}%
  \BibitemOpen
  \bibfield  {author} {\bibinfo {author} {\bibfnamefont {Y.}~\bibnamefont
  {Liu}}, \bibinfo {author} {\bibfnamefont {X.}~\bibnamefont {Yan}}, \bibinfo
  {author} {\bibfnamefont {J.}~\bibnamefont {Wu}}, \bibinfo {author}
  {\bibfnamefont {B.}~\bibnamefont {Zhu}}, \bibinfo {author} {\bibfnamefont
  {Y.}~\bibnamefont {Chen}}, \ and\ \bibinfo {author} {\bibfnamefont
  {X.}~\bibnamefont {Chen}},\ }\bibfield  {title} {\enquote {\bibinfo {title}
  {On-chip erbium-doped lithium niobate microcavity laser},}\ }\href@noop {}
  {\bibfield  {journal} {\bibinfo  {journal} {Science China-Physics Mechanics
  \& Astronomy}\ }\textbf {\bibinfo {volume} {64}},\ \bibinfo {pages} {1--5}
  (\bibinfo {year} {2021})}\BibitemShut {NoStop}%
\bibitem [{\citenamefont {Luo}\ \emph {et~al.}(2021)\citenamefont {Luo},
  \citenamefont {Hao}, \citenamefont {Yang}, \citenamefont {Zhang},
  \citenamefont {Zheng}, \citenamefont {Liu}, \citenamefont {Liu},
  \citenamefont {Bo}, \citenamefont {Kong}, \citenamefont {Zhang} \emph
  {et~al.}}]{ref11}%
  \BibitemOpen
  \bibfield  {author} {\bibinfo {author} {\bibfnamefont {Q.}~\bibnamefont
  {Luo}}, \bibinfo {author} {\bibfnamefont {Z.}~\bibnamefont {Hao}}, \bibinfo
  {author} {\bibfnamefont {C.}~\bibnamefont {Yang}}, \bibinfo {author}
  {\bibfnamefont {R.}~\bibnamefont {Zhang}}, \bibinfo {author} {\bibfnamefont
  {D.}~\bibnamefont {Zheng}}, \bibinfo {author} {\bibfnamefont
  {S.}~\bibnamefont {Liu}}, \bibinfo {author} {\bibfnamefont {H.}~\bibnamefont
  {Liu}}, \bibinfo {author} {\bibfnamefont {F.}~\bibnamefont {Bo}}, \bibinfo
  {author} {\bibfnamefont {Y.}~\bibnamefont {Kong}}, \bibinfo {author}
  {\bibfnamefont {G.}~\bibnamefont {Zhang}},  \emph {et~al.},\ }\bibfield
  {title} {\enquote {\bibinfo {title} {Microdisk lasers on an erbium-doped
  lithium-niobite chip},}\ }\href@noop {} {\bibfield  {journal} {\bibinfo
  {journal} {Science China-Physics Mechanics \& Astronomy}\ }\textbf {\bibinfo
  {volume} {64}},\ \bibinfo {pages} {1--5} (\bibinfo {year}
  {2021})}\BibitemShut {NoStop}%
\bibitem [{\citenamefont {Wang}\ \emph {et~al.}(2021)\citenamefont {Wang},
  \citenamefont {Fang}, \citenamefont {Liu}, \citenamefont {Chu}, \citenamefont
  {Zhou}, \citenamefont {Zhang}, \citenamefont {Wu}, \citenamefont {Wang},
  \citenamefont {Lu},\ and\ \citenamefont {Cheng}}]{ref12}%
  \BibitemOpen
  \bibfield  {author} {\bibinfo {author} {\bibfnamefont {Z.}~\bibnamefont
  {Wang}}, \bibinfo {author} {\bibfnamefont {Z.}~\bibnamefont {Fang}}, \bibinfo
  {author} {\bibfnamefont {Z.}~\bibnamefont {Liu}}, \bibinfo {author}
  {\bibfnamefont {W.}~\bibnamefont {Chu}}, \bibinfo {author} {\bibfnamefont
  {Y.}~\bibnamefont {Zhou}}, \bibinfo {author} {\bibfnamefont {J.}~\bibnamefont
  {Zhang}}, \bibinfo {author} {\bibfnamefont {R.}~\bibnamefont {Wu}}, \bibinfo
  {author} {\bibfnamefont {M.}~\bibnamefont {Wang}}, \bibinfo {author}
  {\bibfnamefont {T.}~\bibnamefont {Lu}}, \ and\ \bibinfo {author}
  {\bibfnamefont {Y.}~\bibnamefont {Cheng}},\ }\bibfield  {title} {\enquote
  {\bibinfo {title} {On-chip tunable microdisk laser fabricated on er3+-doped
  lithium niobate on insulator},}\ }\href@noop {} {\bibfield  {journal}
  {\bibinfo  {journal} {Optics Letters}\ }\textbf {\bibinfo {volume} {46}},\
  \bibinfo {pages} {380--383} (\bibinfo {year} {2021})}\BibitemShut {NoStop}%
\bibitem [{\citenamefont {Chen}\ \emph
  {et~al.}(2021{\natexlab{b}})\citenamefont {Chen}, \citenamefont {Wang},
  \citenamefont {Peng},\ and\ \citenamefont {Yang}}]{chen2021non}%
  \BibitemOpen
  \bibfield  {author} {\bibinfo {author} {\bibfnamefont {W.}~\bibnamefont
  {Chen}}, \bibinfo {author} {\bibfnamefont {C.}~\bibnamefont {Wang}}, \bibinfo
  {author} {\bibfnamefont {B.}~\bibnamefont {Peng}}, \ and\ \bibinfo {author}
  {\bibfnamefont {L.}~\bibnamefont {Yang}},\ }\bibfield  {title} {\enquote
  {\bibinfo {title} {Non-hermitian physics and exceptional points in
  high-quality optical microresonators},}\ }in\ \href@noop {} {\emph {\bibinfo
  {booktitle} {Ultra-high-q Optical Microcavities}}}\ (\bibinfo  {publisher}
  {World Scientific},\ \bibinfo {year} {2021})\ pp.\ \bibinfo {pages}
  {269--313}\BibitemShut {NoStop}%
\bibitem [{\citenamefont {Ma}\ \emph {et~al.}(2011)\citenamefont {Ma},
  \citenamefont {Oulton}, \citenamefont {Sorger}, \citenamefont {Bartal},\ and\
  \citenamefont {Zhang}}]{ref15}%
  \BibitemOpen
  \bibfield  {author} {\bibinfo {author} {\bibfnamefont {R.-M.}\ \bibnamefont
  {Ma}}, \bibinfo {author} {\bibfnamefont {R.~F.}\ \bibnamefont {Oulton}},
  \bibinfo {author} {\bibfnamefont {V.~J.}\ \bibnamefont {Sorger}}, \bibinfo
  {author} {\bibfnamefont {G.}~\bibnamefont {Bartal}}, \ and\ \bibinfo {author}
  {\bibfnamefont {X.}~\bibnamefont {Zhang}},\ }\bibfield  {title} {\enquote
  {\bibinfo {title} {Room-temperature sub-diffraction-limited plasmon laser by
  total internal reflection},}\ }\href@noop {} {\bibfield  {journal} {\bibinfo
  {journal} {Nature Materials}\ }\textbf {\bibinfo {volume} {10}},\ \bibinfo
  {pages} {110--113} (\bibinfo {year} {2011})}\BibitemShut {NoStop}%
\bibitem [{\citenamefont {Nakamura}\ \emph {et~al.}(1975)\citenamefont
  {Nakamura}, \citenamefont {Aiki}, \citenamefont {Umeda},\ and\ \citenamefont
  {Yariv}}]{ref16}%
  \BibitemOpen
  \bibfield  {author} {\bibinfo {author} {\bibfnamefont {M.}~\bibnamefont
  {Nakamura}}, \bibinfo {author} {\bibfnamefont {K.}~\bibnamefont {Aiki}},
  \bibinfo {author} {\bibfnamefont {J.}~\bibnamefont {Umeda}}, \ and\ \bibinfo
  {author} {\bibfnamefont {A.}~\bibnamefont {Yariv}},\ }\bibfield  {title}
  {\enquote {\bibinfo {title} {Cw operation of distributed-feedback gaas-gaalas
  diode lasers at temperatures up to 300 k},}\ }\href@noop {} {\bibfield
  {journal} {\bibinfo  {journal} {Applied Physics Letters}\ }\textbf {\bibinfo
  {volume} {27}},\ \bibinfo {pages} {403--405} (\bibinfo {year}
  {1975})}\BibitemShut {NoStop}%
\bibitem [{\citenamefont {Feng}\ \emph {et~al.}(2014)\citenamefont {Feng},
  \citenamefont {Wong}, \citenamefont {Ma}, \citenamefont {Wang},\ and\
  \citenamefont {Zhang}}]{ref17}%
  \BibitemOpen
  \bibfield  {author} {\bibinfo {author} {\bibfnamefont {L.}~\bibnamefont
  {Feng}}, \bibinfo {author} {\bibfnamefont {Z.~J.}\ \bibnamefont {Wong}},
  \bibinfo {author} {\bibfnamefont {R.-M.}\ \bibnamefont {Ma}}, \bibinfo
  {author} {\bibfnamefont {Y.}~\bibnamefont {Wang}}, \ and\ \bibinfo {author}
  {\bibfnamefont {X.}~\bibnamefont {Zhang}},\ }\bibfield  {title} {\enquote
  {\bibinfo {title} {Single-mode laser by parity-time symmetry breaking},}\
  }\href@noop {} {\bibfield  {journal} {\bibinfo  {journal} {Science}\ }\textbf
  {\bibinfo {volume} {346}},\ \bibinfo {pages} {972--975} (\bibinfo {year}
  {2014})}\BibitemShut {NoStop}%
\bibitem [{\citenamefont {Hodaei}\ \emph {et~al.}(2014)\citenamefont {Hodaei},
  \citenamefont {Miri}, \citenamefont {Heinrich}, \citenamefont
  {Christodoulides},\ and\ \citenamefont {Khajavikhan}}]{ref18}%
  \BibitemOpen
  \bibfield  {author} {\bibinfo {author} {\bibfnamefont {H.}~\bibnamefont
  {Hodaei}}, \bibinfo {author} {\bibfnamefont {M.-A.}\ \bibnamefont {Miri}},
  \bibinfo {author} {\bibfnamefont {M.}~\bibnamefont {Heinrich}}, \bibinfo
  {author} {\bibfnamefont {D.~N.}\ \bibnamefont {Christodoulides}}, \ and\
  \bibinfo {author} {\bibfnamefont {M.}~\bibnamefont {Khajavikhan}},\
  }\bibfield  {title} {\enquote {\bibinfo {title} {Parity-time--symmetric
  microring lasers},}\ }\href@noop {} {\bibfield  {journal} {\bibinfo
  {journal} {Science}\ }\textbf {\bibinfo {volume} {346}},\ \bibinfo {pages}
  {975--978} (\bibinfo {year} {2014})}\BibitemShut {NoStop}%
\bibitem [{\citenamefont {Hodaei}\ \emph {et~al.}(2016)\citenamefont {Hodaei},
  \citenamefont {Miri}, \citenamefont {Hassan}, \citenamefont {Hayenga},
  \citenamefont {Heinrich}, \citenamefont {Christodoulides},\ and\
  \citenamefont {Khajavikhan}}]{ref19}%
  \BibitemOpen
  \bibfield  {author} {\bibinfo {author} {\bibfnamefont {H.}~\bibnamefont
  {Hodaei}}, \bibinfo {author} {\bibfnamefont {M.-A.}\ \bibnamefont {Miri}},
  \bibinfo {author} {\bibfnamefont {A.~U.}\ \bibnamefont {Hassan}}, \bibinfo
  {author} {\bibfnamefont {W.~E.}\ \bibnamefont {Hayenga}}, \bibinfo {author}
  {\bibfnamefont {M.}~\bibnamefont {Heinrich}}, \bibinfo {author}
  {\bibfnamefont {D.~N.}\ \bibnamefont {Christodoulides}}, \ and\ \bibinfo
  {author} {\bibfnamefont {M.}~\bibnamefont {Khajavikhan}},\ }\bibfield
  {title} {\enquote {\bibinfo {title} {Single mode lasing in transversely
  multi-moded pt-symmetric microring resonators},}\ }\href@noop {} {\bibfield
  {journal} {\bibinfo  {journal} {Laser \& Photonics Reviews}\ }\textbf
  {\bibinfo {volume} {10}},\ \bibinfo {pages} {494--499} (\bibinfo {year}
  {2016})}\BibitemShut {NoStop}%
\bibitem [{\citenamefont {Shang}, \citenamefont {Liu},\ and\ \citenamefont
  {Xu}(2008)}]{ref20}%
  \BibitemOpen
  \bibfield  {author} {\bibinfo {author} {\bibfnamefont {L.}~\bibnamefont
  {Shang}}, \bibinfo {author} {\bibfnamefont {L.}~\bibnamefont {Liu}}, \ and\
  \bibinfo {author} {\bibfnamefont {L.}~\bibnamefont {Xu}},\ }\bibfield
  {title} {\enquote {\bibinfo {title} {Single-frequency coupled asymmetric
  microcavity laser},}\ }\href@noop {} {\bibfield  {journal} {\bibinfo
  {journal} {Optics Letters}\ }\textbf {\bibinfo {volume} {33}},\ \bibinfo
  {pages} {1150--1152} (\bibinfo {year} {2008})}\BibitemShut {NoStop}%
\bibitem [{\citenamefont {Wang}\ \emph {et~al.}(2016)\citenamefont {Wang},
  \citenamefont {Xu}, \citenamefont {Jiang}, \citenamefont {Li}, \citenamefont
  {Dai}, \citenamefont {Lu},\ and\ \citenamefont {Li}}]{ref21}%
  \BibitemOpen
  \bibfield  {author} {\bibinfo {author} {\bibfnamefont {Y.}~\bibnamefont
  {Wang}}, \bibinfo {author} {\bibfnamefont {C.}~\bibnamefont {Xu}}, \bibinfo
  {author} {\bibfnamefont {M.}~\bibnamefont {Jiang}}, \bibinfo {author}
  {\bibfnamefont {J.}~\bibnamefont {Li}}, \bibinfo {author} {\bibfnamefont
  {J.}~\bibnamefont {Dai}}, \bibinfo {author} {\bibfnamefont {J.}~\bibnamefont
  {Lu}}, \ and\ \bibinfo {author} {\bibfnamefont {P.}~\bibnamefont {Li}},\
  }\bibfield  {title} {\enquote {\bibinfo {title} {Lasing mode regulation and
  single-mode realization in zno whispering gallery microcavities by the
  vernier effect},}\ }\href@noop {} {\bibfield  {journal} {\bibinfo  {journal}
  {Nanoscale}\ }\textbf {\bibinfo {volume} {8}},\ \bibinfo {pages}
  {16631--16639} (\bibinfo {year} {2016})}\BibitemShut {NoStop}%
\bibitem [{\citenamefont {Gao}\ \emph {et~al.}(2021)\citenamefont {Gao},
  \citenamefont {Guan}, \citenamefont {Yao}, \citenamefont {Deng},
  \citenamefont {Lin}, \citenamefont {Wang}, \citenamefont {Qiao},
  \citenamefont {Wang}, \citenamefont {Liang}, \citenamefont {Zhou} \emph
  {et~al.}}]{gao2021chip}%
  \BibitemOpen
  \bibfield  {author} {\bibinfo {author} {\bibfnamefont {R.}~\bibnamefont
  {Gao}}, \bibinfo {author} {\bibfnamefont {J.}~\bibnamefont {Guan}}, \bibinfo
  {author} {\bibfnamefont {N.}~\bibnamefont {Yao}}, \bibinfo {author}
  {\bibfnamefont {L.}~\bibnamefont {Deng}}, \bibinfo {author} {\bibfnamefont
  {J.}~\bibnamefont {Lin}}, \bibinfo {author} {\bibfnamefont {M.}~\bibnamefont
  {Wang}}, \bibinfo {author} {\bibfnamefont {L.}~\bibnamefont {Qiao}}, \bibinfo
  {author} {\bibfnamefont {Z.}~\bibnamefont {Wang}}, \bibinfo {author}
  {\bibfnamefont {Y.}~\bibnamefont {Liang}}, \bibinfo {author} {\bibfnamefont
  {Y.}~\bibnamefont {Zhou}},  \emph {et~al.},\ }\bibfield  {title} {\enquote
  {\bibinfo {title} {On-chip ultra-narrow-linewidth single-mode microlaser on
  lithium niobate on insulator},}\ }\href@noop {} {\bibfield  {journal}
  {\bibinfo  {journal} {Optics Letters}\ }\textbf {\bibinfo {volume} {46}},\
  \bibinfo {pages} {3131--3134} (\bibinfo {year} {2021})}\BibitemShut {NoStop}%
\bibitem [{\citenamefont {Zhang}\ \emph {et~al.}(2021)\citenamefont {Zhang},
  \citenamefont {Yang}, \citenamefont {Hao}, \citenamefont {Jia}, \citenamefont
  {Luo}, \citenamefont {Zheng}, \citenamefont {Liu}, \citenamefont {Yu},
  \citenamefont {Gao}, \citenamefont {Bo} \emph
  {et~al.}}]{zhang2021integrated}%
  \BibitemOpen
  \bibfield  {author} {\bibinfo {author} {\bibfnamefont {R.}~\bibnamefont
  {Zhang}}, \bibinfo {author} {\bibfnamefont {C.}~\bibnamefont {Yang}},
  \bibinfo {author} {\bibfnamefont {Z.}~\bibnamefont {Hao}}, \bibinfo {author}
  {\bibfnamefont {D.}~\bibnamefont {Jia}}, \bibinfo {author} {\bibfnamefont
  {Q.}~\bibnamefont {Luo}}, \bibinfo {author} {\bibfnamefont {D.}~\bibnamefont
  {Zheng}}, \bibinfo {author} {\bibfnamefont {H.}~\bibnamefont {Liu}}, \bibinfo
  {author} {\bibfnamefont {X.}~\bibnamefont {Yu}}, \bibinfo {author}
  {\bibfnamefont {F.}~\bibnamefont {Gao}}, \bibinfo {author} {\bibfnamefont
  {F.}~\bibnamefont {Bo}},  \emph {et~al.},\ }\bibfield  {title} {\enquote
  {\bibinfo {title} {Integrated lithium niobate single-mode lasers by the
  vernier effect},}\ }\href@noop {} {\bibfield  {journal} {\bibinfo  {journal}
  {Science China Physics, Mechanics \& Astronomy}\ }\textbf {\bibinfo {volume}
  {64}},\ \bibinfo {pages} {1--5} (\bibinfo {year} {2021})}\BibitemShut
  {NoStop}%
\bibitem [{\citenamefont {Xiao}\ \emph {et~al.}(2021)\citenamefont {Xiao},
  \citenamefont {Wu}, \citenamefont {Cai}, \citenamefont {Li},\ and\
  \citenamefont {Chen}}]{xiao2021single}%
  \BibitemOpen
  \bibfield  {author} {\bibinfo {author} {\bibfnamefont {Z.}~\bibnamefont
  {Xiao}}, \bibinfo {author} {\bibfnamefont {K.}~\bibnamefont {Wu}}, \bibinfo
  {author} {\bibfnamefont {M.}~\bibnamefont {Cai}}, \bibinfo {author}
  {\bibfnamefont {T.}~\bibnamefont {Li}}, \ and\ \bibinfo {author}
  {\bibfnamefont {J.}~\bibnamefont {Chen}},\ }\bibfield  {title} {\enquote
  {\bibinfo {title} {Single-frequency integrated laser on erbium-doped lithium
  niobate on insulator},}\ }\href@noop {} {\bibfield  {journal} {\bibinfo
  {journal} {Optics Letters}\ }\textbf {\bibinfo {volume} {46}} (\bibinfo
  {year} {2021})}\BibitemShut {NoStop}%
\bibitem [{\citenamefont {Ge}\ and\ \citenamefont {T{\"u}reci}(2015)}]{ref22}%
  \BibitemOpen
  \bibfield  {author} {\bibinfo {author} {\bibfnamefont {L.}~\bibnamefont
  {Ge}}\ and\ \bibinfo {author} {\bibfnamefont {H.~E.}\ \bibnamefont
  {T{\"u}reci}},\ }\bibfield  {title} {\enquote {\bibinfo {title} {Inverse
  vernier effect in coupled lasers},}\ }\href@noop {} {\bibfield  {journal}
  {\bibinfo  {journal} {Physical Review A}\ }\textbf {\bibinfo {volume} {92}},\
  \bibinfo {pages} {013840} (\bibinfo {year} {2015})}\BibitemShut {NoStop}%
\bibitem [{\citenamefont {Guarino}\ \emph {et~al.}(2007)\citenamefont
  {Guarino}, \citenamefont {Poberaj}, \citenamefont {Rezzonico}, \citenamefont
  {Degl'Innocenti},\ and\ \citenamefont {G{\"u}nter}}]{guarino2007electro}%
  \BibitemOpen
  \bibfield  {author} {\bibinfo {author} {\bibfnamefont {A.}~\bibnamefont
  {Guarino}}, \bibinfo {author} {\bibfnamefont {G.}~\bibnamefont {Poberaj}},
  \bibinfo {author} {\bibfnamefont {D.}~\bibnamefont {Rezzonico}}, \bibinfo
  {author} {\bibfnamefont {R.}~\bibnamefont {Degl'Innocenti}}, \ and\ \bibinfo
  {author} {\bibfnamefont {P.}~\bibnamefont {G{\"u}nter}},\ }\bibfield  {title}
  {\enquote {\bibinfo {title} {Electro--optically tunable microring resonators
  in lithium niobate},}\ }\href@noop {} {\bibfield  {journal} {\bibinfo
  {journal} {Nature photonics}\ }\textbf {\bibinfo {volume} {1}},\ \bibinfo
  {pages} {407--410} (\bibinfo {year} {2007})}\BibitemShut {NoStop}%
\bibitem [{\citenamefont {Hulme}, \citenamefont {Doylend},\ and\ \citenamefont
  {Bowers}(2013)}]{ref23}%
  \BibitemOpen
  \bibfield  {author} {\bibinfo {author} {\bibfnamefont {J.}~\bibnamefont
  {Hulme}}, \bibinfo {author} {\bibfnamefont {J.}~\bibnamefont {Doylend}}, \
  and\ \bibinfo {author} {\bibfnamefont {J.}~\bibnamefont {Bowers}},\
  }\bibfield  {title} {\enquote {\bibinfo {title} {Widely tunable vernier ring
  laser on hybrid silicon},}\ }\href@noop {} {\bibfield  {journal} {\bibinfo
  {journal} {Optics Express}\ }\textbf {\bibinfo {volume} {21}},\ \bibinfo
  {pages} {19718--19722} (\bibinfo {year} {2013})}\BibitemShut {NoStop}%
\end{thebibliography}%

\end{document}